\documentstyle[aps]{revtex}
\input{epsf}
\begin{document}
\flushbottom

\twocolumn

\def\thepage{\roman{page}}
\title{\vspace*{0.5in}Ultra - High Energy Cosmic Rays from decay
 of the Super Heavy Dark Matter Relics.}

\author{A.G. Doroshkevich}

\address{Theoretical Astrophysics Center, Juliane Maries Vej
30, 2100 Copenhagen, {\O} Denmark;}

\author{P.D. Naselsky}

\address{Theoretical Astrophysics Center, Juliane Maries Vej
30, 2100 Copenhagen, {\O} Denmark;\\ Rostov State University, 
Zorge 5, 344090, Rostov-Don, Russia}

\maketitle
\newcommand{\bedm}{\begin{displaymath}}
\newcommand{\eedm}{\end{displaymath}}
\newcommand{\be}{\begin{equation}}
\newcommand{\ee}{\end{equation}} 
\newcommand{\g}{\nabla}
\newcommand{\de}{\partial}    
\newcommand{\ha}{\frac{1}{2}}
\newcommand{\ci}[1]{\cite{#1}} 
\newcommand{\bi}[1]{\bibitem{#1}}
\newcommand{\noi}{\noindent}

\newcommand{\ga}{\alpha}
\newcommand{\gb}{\beta}
\newcommand{\gc}{\gamma}
\newcommand{\gd}{\delta}
\newcommand{\gep}{\epsilon}
\newcommand{\gee}{\varepsilon}
\newcommand{\gz}{\zeta}
\newcommand{\get}{\eta}
\newcommand{\gth}{\theta}
\newcommand{\gthh}{\vartheta}
\newcommand{\gi}{\iota}
\newcommand{\gk}{\kappa}
\newcommand{\gl}{\lambda}
\newcommand{\gm}{\mu}
\newcommand{\gn}{\nu}
\newcommand{\gks}{\xi}
\newcommand{\go}{\0}
\newcommand{\gp}{\pi}
\newcommand{\gpp}{\varpi}
\newcommand{\gr}{\rho}
\newcommand{\grr}{\varrho}
\newcommand{\gs}{\sigma}
\newcommand{\gss}{\varsigma}
\newcommand{\gt}{\tau}
\newcommand{\gu}{\upsilon}
\newcommand{\gf}{\varphi}
\newcommand{\gff}{\varphi}
\newcommand{\gx}{\chi}
\newcommand{\gps}{\psi}
\newcommand{\gw}{\omega}
\newcommand{\gG}{\Gamma}
\newcommand{\gD}{\Delta}
\newcommand{\gTh}{\Theta}
\newcommand{\gL}{\Lambda}
\newcommand{\gKs}{\Xi}
\newcommand{\gP}{\Pi}
\newcommand{\gS}{\Sigma}
\newcommand{\gU}{\Upsilon}
\newcommand{\gF}{\phi}
\newcommand{\gPs}{\Psi}
\newcommand{\gW}{\Omega}

\newcommand{\ti}{\tilde}
\newcommand{\Li}{{\cal L}}
\newcommand{\ra}{\rightarrow}
\newcommand{\pa}{\partial}
\newcommand{\ov}{\overline}
\newcommand{\fad}{\frac{\Delta T}{T}}
\newcommand{\lan}{\langle}
\newcommand{\ran}{\rangle}
\flushbottom

\begin{abstract}
In this paper we briefly discuss  the problem of the origin of Ultra High 
Energy Cosmic Rays  in the framework of Top-Down models. We show that, 
for high energy of decays and in a wide range of spectra of injected protons, 
their extragalactic flux is  consistent with the observed fluxes of cosmic 
rays in the energy range $0.1 E_{GZK}\leq E\leq 10E_{GZK}$. For suitable 
energy and spectra of injected protons, the contribution of galactic sources 
is moderate, in this energy range, but it dominates at smaller and 
larger energies. In such models we can expect that at these energies the 
anisotropy of cosmic rays distribution over sky will be especially small. 

Some possible manifestations of decays of super massive particles such as, 
for example, primordial black holes with masses $M_{pbh}\sim 10^{-5} g$, 
are considered. In particular, we show that partial conversion of energy 
released during these decays at redshifts $z\sim$ 1000 to Ly$-\alpha$ 
photons can delay the hydrogen recombination and distort the spectrum of 
fluctuations of the cosmic microwave background radiation.

PACS number(s): 98.80.Cq, 95.35.+d, 97.60.Lf, 98.70.Vc.
\end{abstract}

\section {Introduction}

The origin of Ultra -- High Energy Cosmic Rays (UHECR) with 
energy above the Greisen-Zatsepin-Kuzmin (GZK) \cite{1,2} cutoff, 
$E_{GZK}\sim 10^{20}$ eV, is one of the most intriguing mysteries 
of the modern physics and astrophysics. After the pioneering papers 
\cite{1,2} and recent observations of the UHECR energy power spectra 
by AGASA \cite{3}, Fly's Eye \cite{4} and Haverah Park \cite{5}, 
several possible mechanisms of the UHECR production were discussed 
(see reviews \cite{6,7}). 

In this paper we consider the so-called Top-Down scenario of the UHECR 
creation which is associated with decays of Super Heavy Dark 
Matter (SHDM) particles with masses $m_{SHDM}>10^{12}$ GeV. This 
mechanism was for the first time suggested by Berezinsky, Kachelrie$\ss$ and 
Vilenkin \cite{8} (see also \cite{9}) and now it seems to be a 
natural way of explaining the origin of UHECR with energies 
above the GZK cutoff.

Several kinds of SHDM particles which could be created in 
the early Universe are discussed in literature. Particles 
with masses of about one to two orders of magnitude larger than the 
typical mass of inflaton, $m_\gF\sim 10^{13}$ GeV, could be very 
efficiently created at the preheating phase of inflation \cite{11}. 
SHDM particles can also be related to  topological defects, 
such as strings \cite{12,13}, magnetic monopoles \cite{14,15}, 
necklaces \cite{16} and vortons \cite{17}. 

The possible contribution of  primordial black hole relics 
(PBHs) to the dark matter (DM) were discussed already in \cite{mg} 
and \cite{br}. Recently Dolgov, Naselsky and Novikov \cite{20} 
considered PBHs with masses $M_{PBH}\sim 10^6$ g as possible sources 
of the baryonic asymmetry and the high entropy of the Universe. 
They assume that remnants of such black holes with masses of about 
Planck mass survive up to now and form the SHDM relics. Mergers of 
these remnants within high density clumps creates more massive black 
holes, stimulates their explosive evaporation and could  produce 
ultra--high energy particles observed as rare UHECRs. 

This discussion shows that, in the framework of the Top-Down 
scenario, the UHECR can be related to various kinds of SHDM 
particles with masses $10^{12}\leq m_X\leq 10^{19}$GeV. (Below by  
$X$ we denote  all possible types of SHDM).

Now the possible energy losses of the UHECR are well established 
\cite{21} and observational predictions of the Top-Down scenario 
of the UHECR creation crucially depend upon unknown factors such 
as the mass of the SHDM particles, energy spectrum and composition 
of decay products. It is commonly believed, that the observed UHECR 
spectrum at both $E\leq E_{GZK}$ and $E\geq E_{GZK}$ is dominated 
mainly by local sources and it simply reproduces the spectra of injected 
protons. However, at energies $E\sim E_{GZK}$, the complex shape of 
observed UHECR fluxes shows that it can be more sensitive to 
extragalactic component of high energy protons and, so, can depend 
upon the unknown factors mentioned above. Detailed investigation of 
the spectrum and anisotropy of UHECR in the range $E/E_{GZK}\geq$ 0.1 
can discriminate between discussed versions of the Top-Down models 
and restrict some parameters of the SHDM particles and the 
process of proton creation.  

As is commonly believed, decays of SHDM particles into the high 
energy protons, photons, electron-positron pairs and neutrinos 
occurs through the production of quark-antiquark pairs ($X\ra 
{q},\ov{q}$), which rapidly hadronize and generate two jets and 
transform the energy into pions ($\sim$95\%) and hadrons ($\sim$
5\%) \cite{21}. It can be expected that later most of that 
energy is transformed into high energy photons and neutrinos 
with the energy spectrum $S(E)\propto E^{-1.5}$, at $E\ll M_X$ 
\cite{21}. Similar spectrum for the hadronic component is also 
expected. This means that, for such decays of SHDM particles with 
$10^{12}<{m_X}<10^{19}$ GeV, the UHECR with energies $E>10^{20}$ eV 
are dominated by photons and neutrinos \cite{21}. This conclusion 
can be tested with further observations of the UHECR fluxes at 
$E>10^{20}$eV. 

Other spectra of protons generated by decays of SHDM particles 
are also discussed. In particular, such spectrum can be similar 
to Gaussian or $\gd$-function centered at $E_X\sim m_X\gg 10^{20}$eV. 
In this case the spectrum of protons created by nearby sources will 
be also similar to the same $\gd$-function while the spectrum of 
the extragalactic component is $\propto E^{-1}$ at both $E< E_{GZK}$ 
and $E> E_{GZK}$ and is  consistent with the observed one at 
$E\sim E_{GZK}$. 

Recently Berezinsky and Kachelrie$\ss$ \cite{13} discussed Monte 
Carlo simulations of the jet fragmentation in SUSY- QCD. They 
found that the spectrum of injected protons can be well fitted to 
log-normal distribution. Farrar and Piran \cite{23} discussed the 
spectrum of injected protons $S(E)\propto E^{-\alpha}$, with 
$0< \alpha\leq$ 1 in the Top-Down model of the UHECR origin. We show 
that, for such spectra, the flux of extragalactic protons at $E\sim 
E_{GZK}$ is also consistent with the observed one. 

Another important factor is the observed anisotropy of the UHECR 
distribution over the sky. As was discussed by Berezinsky et al. 
\cite{8}, both CDM and SHDM particles are clustered within galactic 
halos and their decays inevitably generate some anisotropy. In 
particular, if the contribution of Galactic sources dominates we 
will see an anisotropy of the UHECR due to our asymmetric position 
in the Galaxy. In contrast, the extragalactic component of UHECRs 
is averaged over the volume with a size $\geq$ 50 Mpc and its angular 
distribution is almost isotropic. The contributions of closest 
galaxies and the Local Supercluster of galaxies could also be observed. 

The relative contributions of Galactic and extragalactic UHECRs sources 
depend upon many unknown factors such as the size of galactic halo, overdensity 
and spatial distribution of SHDM particles within the halo \cite{6}. However, 
for larger energy of injection, $E_{inj}\geq 10^4 - 10^5 E_{GZK}$, 
and for Gaussian, log-normal and power spectra of injected protons 
with $\alpha\leq$ 0.6, the contribution of extragalactic UHECRs dominates 
at $E\sim E_{GZK}$, whereas at both less and larger energies the 
contribution of galactic sources becomes more important. This means 
that the anisotropy of angular distribution of UHECRs depends upon 
their energy and, for the Top-Down model with spectra under discussion, 
it is minimal at $E\sim E_{GZK}$. 

In this paper we recalculate the contribution of the extragalactic
component of UHECR for different life -- time, masses of the SHDM 
particles $10^{12}$GeV$\leq{m_X}\leq10^{19}$GeV, and spectra of 
injected proton. We show that the relative contributions of galactic 
and extragalactic components of high energy protons depend upon these 
factors. For the most interesting models, the extragalactic component 
is found to be dominant at $E\sim E_{GZK}$ and the expected flux is well 
consistent with observations. The expected angular distribution on the 
sky of observed UHECR flux depends upon the energy of protons near the 
GZK cutoff and its variations should be considered as an important 
test for the Top--Down models. 

Some cosmological manifestations of decays of SHDM particles can 
also be observed. In particular, decays of SHDM particles with 
masses $m_X\sim 10^{19}$ GeV can delay the recombination of hydrogen 
at redshifts $z\sim$ 1000. This inference is especially important for 
discussion of the Cosmic Microwave Background (CMB) anisotropy and 
polarization power spectra. The same decays can increase the hydrogen 
ionization at smaller redshifts and accelerate the formation of first 
population of stars and galaxies. These manifestations can be tested 
with both available and future measurements of the CMB anisotropy 
and polarization. 

The  paper is organized as  follows. In section 2 we discuss 
the spectra of extragalactic protons for different spectra of injected 
protons, in section 3 the combined flux of galactic and extragalactic 
sources is compared with observations. In section 4 we discuss the 
possible delay of hydrogen recombination due to decays of the 
SHDM particles. Main results are discussed in section 4.
 
\section{Expected spectrum of high energy protons}

In this paper we use the continual energy loss (CEL) approximation 
and describe the evolution of the number density of high energy 
extragalactic protons by the following equation\cite{6}:
\be
{\partial N(E,t)\over \partial t}+3H(t)N+{\partial\over \partial E}
\left(N {dE\over dt}\right) = I(E,t),\\
\label{eq1}
\ee
\be
{1\over E}{dE\over dt} = - H+H_0\beta(E),\quad
 \beta(E)= \beta_\gamma(E)+\beta_\pi(E),
\label{eq2}
\ee
\bedm
I = \gw_p{n_0\over\tau_0}(1+z)^3 S(E),~
H = -{1\over 1+z}{dz\over dt} = H_0(1+z)^{3/2},
\eedm
where $E$ and $N(E,t)$ are the energy and number density of protons per 
unit of energy, $H(t)$ is the Hubble constant, $H_0=100~h$ km/s/Mpc = 
75 km/s/Mpc, $z$ is a redshift, the functions $I(E,t)$ and $S(E)$ 
characterize the intensity and spectrum of injected protons, 
$\gw_pn_0/\tau_0$ is the intensity of proton production at $z=0$ , 
functions $\beta_\gamma(E)$ and $\beta_\pi(E)$ describe the proton 
energy losses due to electron -- positron pairs and photo-pions 
production and $\gw_p\sim 0.05$ is the fraction of protons in the SHDM 
particles decay. 

The general solution of equation (\ref{eq1}) is 
\bedm
N_{ex}(E_0,t_0) = \int_1^\infty {dx\over x^4}{E\over E_0}
{x^{-1}+\beta(E)\over 1+\beta(E_0)}{I(E,x)\over H(x)} = 
\eedm
\be
{\gw_pn_0\over H_0\tau_0}\int_1^\infty {dx\over x^{5/2}}{E\over E_0}{x^{-1}+
\beta(E)\over 1+\beta(E_0)}S(E(x)),
\label{ne}
\ee
where $E=E(x)$, $E_0=E(z=0)$, $t_0=t(z=0)$. This means that, in fact, 
the observed flux of extragalactic UHECRs depends upon the functions 
$\beta(E)$ and $S(E)$.
 
\begin{figure}
\centering
\epsfxsize=8 cm
\epsfbox{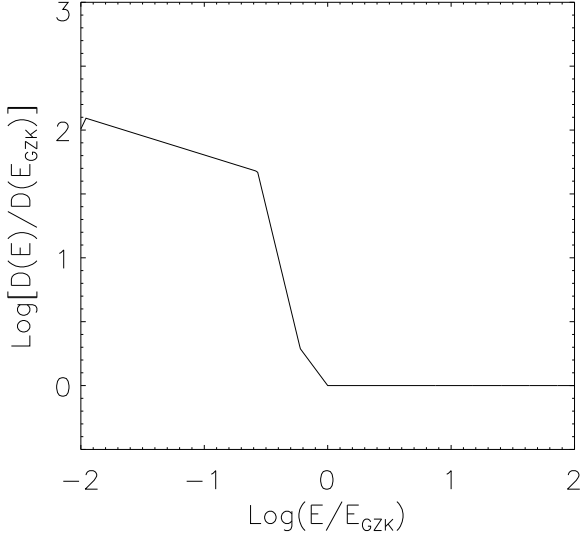}
\vspace{0.75cm}
\caption{The function $D(E)/D(E_{GZK})$ versus $\epsilon=E/E_{GZK}$ for 
photo-pion production.} 
\end{figure}

\subsection{Energy losses of protons}

For protons with $E\geq E_{GZK}(1+z)^{-1}$, the free path, $D(E)$, 
is determined by photo-pion production on CMB photons. Following 
\cite{21} we approximate this energy loss by the function:
\begin{center}
\be  
\beta_\pi=(1+z)^3\kappa_\pi,~~ (1+z)\epsilon\geq 1,
\ee
\be
\beta_\pi = (1+z)^3\kappa_\pi \epsilon^{p_1},~~  1\geq (1+z)\epsilon\geq
\epsilon_1,~~~~~~~~
\ee
\be
\beta_\pi = (1+z)^3\kappa_\pi \epsilon^{p_2}\epsilon_1^{p_1-p_2},~~ 
\epsilon_1 \geq  (1+z)\epsilon\geq\epsilon_2,
\ee
\be
\beta_\pi = (1+z)^3\kappa_\pi\epsilon^{p_3}\epsilon_1^{p_1-p_2}
\epsilon_2^{p_2-p_3},~ \epsilon_2 \geq  (1+z)\epsilon,
\ee
\bedm
\epsilon_1\approx 0.6, ~~\epsilon_2\approx 0.27,~ p_1\approx 1.3,
~ p_2\approx 4,~ p_3\approx 0.3,
\eedm
\bedm
\epsilon=E/E_{GZK},~~E_{GZK}\approx 1.5\cdot 10^{20} eV,
\eedm
\bedm
\kappa_\pi={cH_0^{-1}\over D(E_{GZK})}\approx 160,~~
D(E_{GZK})=25 {\rm Mpc}.
\label{beta_pi}
\eedm
\end{center}
where the redshift dependence of density and temperature of CMB is 
taken into account. The function $\beta(E_{GZK})/\beta(E) = 
D(E)/D(E_{GZK})$ is plotted in Fig. 1. 

For $E\ll E_{GZK}$, the energy losses due to $e^+e^-$  pair production 
dominates (for review, see \cite{21}) and 
\be
\beta_\gamma\approx 0.005(1+z)^3\kappa_\pi. 
\label{beta_ee}
\ee

For these functions $\beta(E)$, the redshifts variations of 
the proton energy can be found analytically as follows: 
\be
E(z_2) = E(z_1){1+z_2\over 1+z_1}G_e(z_1,z_2),~ \beta(E)=\kappa=const,
\label{e_pi}
\ee
\bedm
G_e(z_1,z_2)=\exp\left({2\over 3}\kappa[(1+z_2)^{3/2}-(1+z_1)^{3/2}
]\right), 
\eedm
\be
E(z_2) = E(z_1){1+z_2\over 1+z_1}G_p(z_1,z_2),~ 
\beta(E)=\kappa~\epsilon^p,
\label{e_p}
\ee
\bedm
G_p^{-p}(z_1,z_2)=1-\kappa{p\epsilon^p(z_1)\over p+3/2}
(1+z_1)^{3/2}\left[\left({1+z_2\over 1+z_1}\right)^{p+3/2}
-1\right].
\eedm
 For $p\rightarrow 0$, $G_p(z_1,z_2)\rightarrow G_e(z_1,z_2)$ and 
the expression (\ref{e_p}) becomes identical to (\ref{e_pi}). 

\subsection{Galactic and extragalactic  protons}

The high concentration of the SHDM particles within the halo of 
Galaxy generates the galactic component of UHECRs, $N_{gal}$. 
Its spectrum  reproduces the spectrum of generated protons. 
For the popular King's profile of density distribution in halo, 
\be
\rho(r)=\rho_c(1+r^2/r_g^2)^{-3/2},
\label{king}
\ee
with the size of the core $r_g\sim 10$kpc (see, e.g., \cite{24}), 
the relative contribution of galactic and extragalactic components 
can be roughly estimated as follows: 
\be
N_{gal}(E) = \gw_p{n_0\over\tau_0}{r_g\over c}\delta_gS(E)=
\zeta_{gal}{\gw_pn_0\over H_0\tau_0}S(E),
\ee
\be
\zeta_{gal}\sim {H_0r_g\over c}\delta_g=0.3 {r_g\over 10{\rm kpc}}
{\delta_g\over 10^5}.
\label{gal}
\ee
where $\delta_g$ is the overdensity of the core above the mean density 
of DM component. 
More detailed analysis \cite{6} extends the range of possible ratio 
of galactic to extragalactic components up to $\zeta_{gal}\sim$30 -- 
50. Further on for comparison of galactic and extragalactic fluxes, 
we will take $\zeta_{gal}\sim$ 10. 

\subsubsection{Spectra of extragalactic  protons}

Both the observed fluxes and relative contributions of galactic and 
extragalactic components depend upon the spectrum of injected protons, 
$S(E)$. To illustrate this dependence we consider the normalized power 
spectra with exponents $\alpha\geq$ 1 and $\alpha\leq$ 1 and the 
log-normal spectrum proposed in \cite{13}.
 
Spectrum with $\alpha = 1.5>1$ and $E_{min}\ll E\leq E_{inj}$,
\be
S_{pw}(E)={\alpha-1\over E_{min}}\left({E\over E_{min}}
\right)^{-\alpha}\left(1-{E\over E_{inj}}\right)^2,
\label{spc1}
\ee
is usually used to describe the decay of particles with moderate 
masses. It is model dependent and its applicability to the decay 
of extremely massive X-particles is in question (see, e.g., 
discussion in \cite{13}). For such spectrum, using (\ref{ne}), we obtain
\be
N_{ex} = {\gw_pn_0\over H_0\tau_0}{\alpha-1\over E_{min}}\left({E_0
\over E_{min}}\right)^{-\alpha}\nu(E_0,\alpha).
\label{n15}
\ee
The dimensionless function $\nu(E_0,\alpha)$ weakly depends upon $\alpha$ 
and describes how the flux of extragalactic protons varies 
with energy at $E_0\sim E_{GZK}$. Numerically, $\nu(E_{GZK},1.5)
\approx 10^{-2}$. 

As is seen from (\ref{n15}) in this case both spectra of extragalactic 
and galactic protons are similar and the contribution of extragalactic  
component to the observed fluxes of UHECR is small, because $\nu\ll\zeta_{gal}$.

Spectra of injected protons with $\alpha\leq 1$,  
\be
S_{pw}(E)={c(\alpha)\over E_{inj}}
\left(E\over E_{inj}\right)^{-\alpha}\left(1-{E\over E_{inj}}\right)^2,
\label{spc05}
\ee
\bedm
c(\alpha)=0.5(1-\alpha)(2-\alpha)(3-\alpha),
\eedm
seem to be more promising. For such spectra with $\alpha\leq$ 0.6 
and $E_{inj}\geq 10^2 E_{GZK}$, we have almost universal spectrum 
of extragalactic protons,
\be
N_{ex}(E_0) = {\gw_pn_0\over H_0\tau_0} E_0^{-1}\mu(E_0,E_{inj},\alpha),
\label{n05}
\ee
\be
\mu(E_{GZK},E_{inj},\alpha)\approx (0.5 - 1.5)\cdot 10^{-2}.
\ee
for models with large and short life -- time of the SHDM particles, 
respectively. Dimensionless function $\mu(E_0,\alpha)$ weakly depends 
upon $E_{inj}$ and $\alpha$ and describes variations of the flux of 
extragalactic protons with energy at $E_0\sim E_{GZK}$. 
For such spectra of injected protons galactic component dominates only 
at high energy when 
\be
\left({E_0\over E_{inj}}\right)^{1-\alpha}\geq {\mu(E_0,E_{inj},\alpha)\over 
c(\alpha)\zeta_{gal} }.
\label{dom05}
\ee

We consider also the log-normal spectra of injected protons recently 
proposed in \cite{13} ( see also \cite{6}) ,
\be
S_{ln}(E)={1\over\sqrt{2\pi}E\sigma_{inj}}\exp\left(-{\ln^2(E/E_{inj})
\over 2\sigma_{inj}^2}\right),
\label{sgs}
\ee
with $\sigma_{inj}=$3 -- 7 and the same two energies of injection as above. 
In this case the spectrum of extragalactic protons is also almost 
universal and similar to (\ref{n05}) with a similar function 
$\mu(E_0,E_{inj},\sigma_{inj})$. For such spectrum, the contribution 
of galactic sources is shifted to energy $E_0\sim E_{inj}\exp(-
\sigma_{inj})$
and dominates only at high energy when
\be
\ln^2\left(E_0\over E_{inj}\right)\leq 2\sigma_{inj}^2\ln\left(
\zeta_{gal}\over\sqrt{2\pi}\mu\sigma_{inj}\right).
\label{lgl}
\ee

\subsubsection{Cumulative fluxes of galactic and extragalactic protons}

As is seen from (\ref{n15}) for the spectra of injected protons with 
$\alpha\geq$ 1 the contribution of galactic protons dominates at all 
energies for $\zeta_{gal}\geq \nu(E_{GZK},\alpha)\sim 10^{-2}$. In 
contrast, for both spectra with power index $\alpha\leq$ 1 and log--
normal spectra of injected protons, the galactic component dominates 
only at higher energies as is given by (\ref{dom05}) and (\ref{lgl}). 

Comparison of the {\it cumulative} fluxes of these components shows  
that, for both spectra (\ref{spc05}) and (\ref{sgs}), the extragalactic 
component dominates when 
\be
\zeta_{gal}\leq \mu(E_{GZK})\ln(E_{inj}/E_{GZK}). 
\label{cum}
\ee
Due to the universal spectrum of extragalactic protons (\ref{n05}), 
this estimate does not depend upon detailed characteristics 
of spectra of injected protons and shows that even for $E_{inj}\sim 
10^8 E_{GZK}$ the cumulative extragalactic component dominates only 
for $\zeta_{gal}\leq$ 0.1 -- 0.3, for large and short life -- time 
of the SHDM particles. These values are close to the estimate 
(\ref{gal}) but are less then those discussed in \cite{6}. This means 
that, for both spectra (\ref{spc05}) and (\ref{sgs}), the domination 
of galactic component could be expected at $E\sim E_{inj}\gg E_{GZK}$. 

\section{Expected flux of protons for Top--Down models}

Here we consider the Top--Down models for two injection energy, 
$E_{inj}=10^2E_{GZK}$ and $E_{inj}=10^8E_{GZK}$. The first model is 
related to decays of SHDM particles with moderate masses often 
discussed in literature (see, e.g., \cite{6,7}). The second model 
describes the decay of ultra massive particles such as, for 
example, explosive evaporation of black hole remnants discussed 
in Dolgov, Naselsky \& Novikov \cite{20}. 

To illustrate the influence of the life -- time of the SHDM particles, 
$\tau_0$, we consider two models, one with $\tau_0$ larger than the age 
of the Universe, $T_U\sim H_0^{-1}$, and other with $\tau_0\approx 
0.1 T_U$.

The normalized expected fluxes of UHE protons, 
\be
F(\epsilon_0)={dJ(E_0)\over dE_0}{E_0^3\over 10^{24}eV^2 m^{-2} 
ster^{-1}s^{-1}}
\label{fe}
\ee
are plotted in Figs. 2 -- 5 versus $\epsilon_0 = E_0/E_{GZK}$ 
together with available observational data. We show 
the fluxes of extragalactic component alone and combined fluxes 
for extragalactic and galactic components for $\zeta_{gal}$=10. 

\begin{figure}
\centering
\epsfxsize=8 cm
\epsfbox{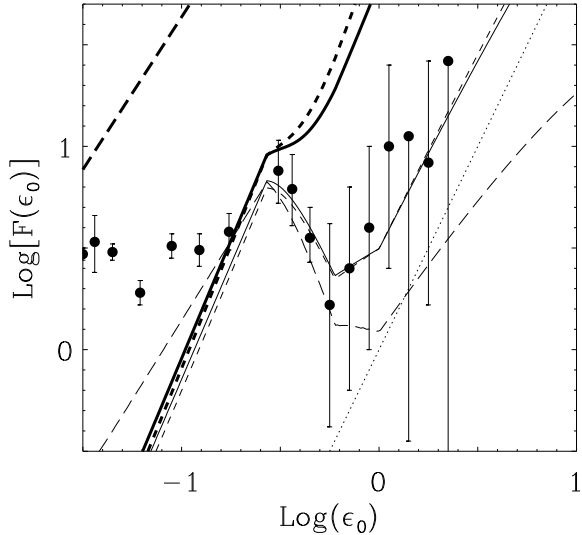}
\vspace{0.75cm}
\caption{The functions $F(\epsilon_0)$ (\ref{fe}) versus $\epsilon_0=
E/E_{GZK}$ are plotted for long-lived SHDM particles with $\tau_0\gg 
T_U$, $E_{inj}/E_{GZK}=10^2$, and power spectra of injected 
protons with $\alpha=1.5$ (long dashed line), $\alpha=0.5$ (solid line), 
and $\alpha=0.25$ (dashed line). Thin lines show the contribution 
of extragalactic component alone, thick lines show the contribution 
of extragalactic and galactic sources for $\zeta_{gal}=10$.
The observed fluxes  are  plotted by points. For comparison, the flux 
$dJ/dE_0\propto E_0^{-1}$ is plotted by dotted line. 
}
\end{figure}

\subsection{Models with power spectra of injected protons} 

As was noted above, for power spectra (\ref{spc1}) with larger exponent 
$\alpha=1.5\geq 1$, the contribution of extragalactic component, 
$N_{ex}(E_0)$, weakly depends upon the energy of injection and is 
negligible in comparison with the contribution of the galactic component. 
These results are a natural consequence of predominant generation of 
lower energy protons in such models. Of course, for suitable choice of 
decay rate, $n_0/\tau_0$, the galactic component can explain the 
observed growth of flux at $E\geq E_{GZK}$. 

\subsubsection{Models with $\alpha\leq$ 1 and larger life--time 
of the SHDM particles, $\tau_0\gg T_U$} 

For models with smaller exponents, $\alpha=$0.5 and 0.25, and moderate 
energy of injection, $E_{inj}=10^2E_{GZK}$, plotted in Fig. 2, the
resulting flux is sensitive to the contribution of galactic component 
and, for the most interesting energies $E_0\geq 0.2 - 0.3E_{GZK}$, 
the impact of extragalactic component becomes noticeable only for 
$\zeta_{loc}\leq$ 1. These results agree with approximate 
estimates (\ref{dom05}).

For models with high energy of injection, $E_{inj}=10^8E_{GZK}$, 
and with smaller exponents, $\alpha=$0.5 and 0.25, the expected spectrum 
of extragalactic component is similar to (\ref{n05}), and the resulting 
flux is weakly sensitive to the contribution of the galactic
component.  For the most interesting 
energies, $E_0\sim E_{GZK}$, the extragalactic component dominates, 
at least for $\zeta_{gal}\leq$30. 

Results plotted in Fig. 3 show that, for suitable decay rate,
\be
\gw_p n_0/\tau_0\approx 10^{-46}cm^{-3}s^{-1},
\label{ntl}
\ee
such models reproduce quite well the observed fluxes of UHECR with energies 
$E\geq 0.1 E_{GZK}$. 

\begin{figure}
\centering
\epsfxsize=8 cm
\epsfbox{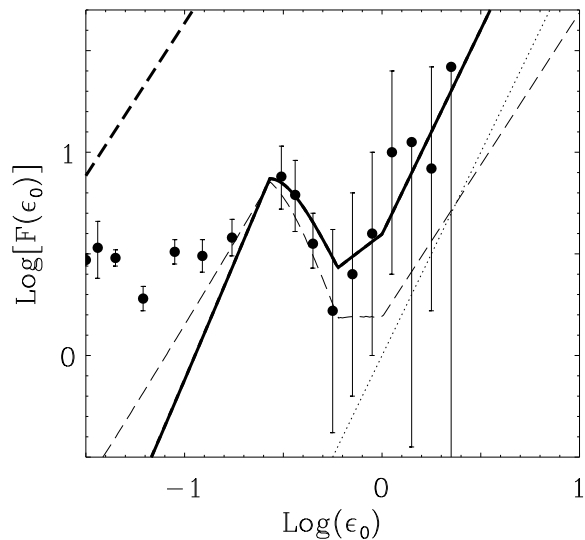}
\vspace{0.75cm}
\caption{The functions $F(\epsilon_0)$ (\ref{fe}) versus $\epsilon_0=
E/E_{GZK}$ are plotted for long-lived SHDM particles with $\tau_0\gg 
T_U$, $E_{inj}/E_{GZK}=10^8$, and power spectra of injected 
protons with $\alpha=1.5$ (long dashed line), $\alpha=0.5$ 
 (solid line). 
 Thin lines show the contribution 
of extragalactic component alone, thick lines show the contribution 
of extragalactic and galactic sources, for $\zeta_{gal}=10$. 
The observed fluxes are plotted by points. For comparison, the flux 
$dJ/dE_0\propto E_0^{-1}$ is plotted by dotted line. 
}
\end{figure}

\subsubsection{Models with shorter life-time of the SHDM particles} 

In models with shorter life -- time of the SHDM particles, $\tau_0\sim 
0.125 T_U$, the number density of  SHDM particles rapidly decreases 
with time due to their progressive decays what, in turn, increases the 
contribution of extragalactic component for  
0.1 $\leq E_0/E_{GZK}\leq$ 1. 
In spite of this, for models with moderate energy of injection, 
$E_{inj}=10^2E_{GZK}$, this factor cannot essentially 
amplify the contribution of extragalactic component 
and it becomes noticeable only for $\zeta_{gal}<$10.

However, for models with ultra -- high energy of injection, 
$E_{inj}=10^8E_{GZK}$, and smaller exponents, $\alpha=0.5$ and 
$\alpha=0.25$, results plotted in Fig. 4 for $\tau_0=0.125 T_U$, 
demonstrate that, for $E_0\geq 0.06E_{GZK}$, the extragalactic
component dominates. For the decay rate
 \be
{\gw_pn_0\over\tau_0}\approx 0.3\cdot10^{-46}cm^{-3}s^{-1},~ 
n_0\sim \gw_p^{-1} 10^{-30}cm^{-3},
\label{nts}
\ee
it reproduces quite  well the observed flux of UHECR for energy $E\geq 
0.06 E_{GZK}$. If  SHDMs are identified with primordial 
black holes with $M_{pbh}\sim 10^{-5} g$ then the mean densities 
of SHDMs at $z=0$ and at $z\gg$ 1 are
\be
\rho(0)\sim \gw_p^{-1} 10^{-35}g~cm^{-3},
\ee
\be
{\rho(z)\over (1+z)^3}\sim \gw_p^{-1}\left(10^{-31}-
10^{-32}\right) g~cm^{-3}\leq \rho_{cr},
\label{rho}
\ee
respectively.

\begin{figure}
\centering
\epsfxsize=8 cm
\epsfbox{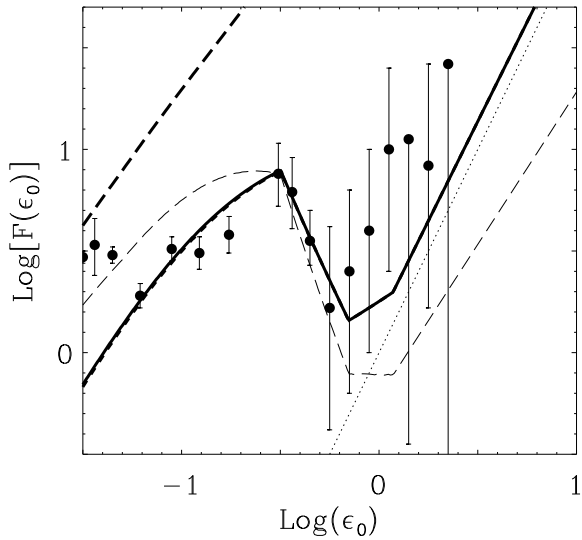}
\vspace{0.75cm}
\caption{The functions $F(\epsilon_0)$ (\ref{fe}) versus $\epsilon_0=
E/E_{GZK}$ are plotted for short-lived SHDM particles with $\tau_0= 
0.125 T_U$, $E_{inj}/E_{GZK}=10^8$, and power spectra of injected 
protons with $\alpha=1.5$ (long dashed line), $\alpha=0.5$ (solid line). 
 Thin lines show the contribution 
of extragalactic component alone, thick lines show the contribution 
of extragalactic and galactic sources, for $\zeta_{gal}=10$. 
The observed fluxes are plotted by points. For comparison, the flux 
$dJ/dE_0\propto E_0^{-1}$ is plotted by dotted line. 
}
\end{figure}

\subsection{Models with log-normal spectra of injected protons} 

For log-normal spectra of injected protons, $S_{ln}$, (\ref{sgs}), 
with  moderate dispersions $\sigma_{inj}=5$ and energies of 
decays $E_{inj}=10^2 E_{GZK}$ and $E_{inj}=10^8 E_{GZK}$, the resulting 
fluxes of UHECRs, for $\zeta_{gal}=10$,  are plotted in Fig. 5. For 
models with larger $E_{inj}$, the expected flux at $E_0\sim E_{GZK}$ 
is dominated by the extragalactic component and reproduces the 
observed flux variations. In contrast, for models with smaller $E_{inj}$ 
the expected flux is dominated by the galactic component and is far 
from the observed one. The contribution of galactic sources at 
$E_0\sim E_{GZK}$ depends upon $\zeta_{gal}$ and rapidly decreases 
for larger $E_{inj}$ and smaller $\sigma_{inj}$.

At $E\leq E_{GZK}$ the extragalactic flux is more sensitive to the 
life -- time of SHDM particles. As is seen from Fig. 5 for the 
decay rate of SHDM particles (\ref{ntl}) and longer life -- 
time, $\tau_0\gg T_U$, the resulting fluxes, for 
$E_0\geq 0.3 E_{GZK}$,
describe quite well the observations. For shorter life -- time, 
$\tau_0\sim 0.1T_U$, and the decay rate of SHDM particles 
(\ref{nts}), this fluxes reproduce well the observations up to 
$E_0\sim 0.06 E_{GZK}$. 

\begin{figure}
\centering
\epsfxsize=8 cm
\epsfbox{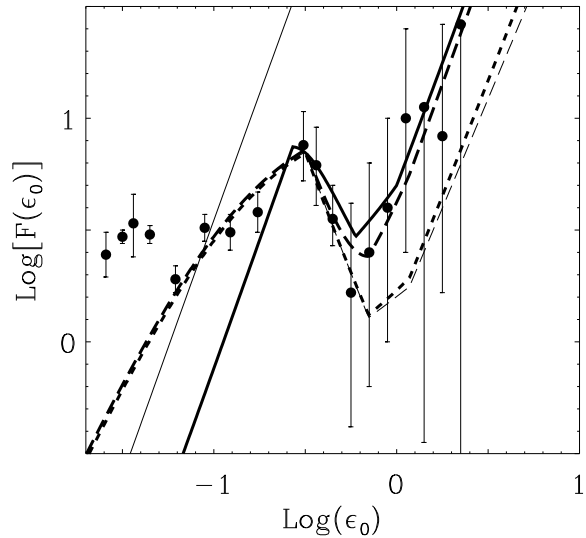}
\vspace{0.75cm}
\caption{The functions $F(\epsilon_0)$ (\ref{fe}) versus $\epsilon_0=
E/E_{GZK}$ are plotted for log-normal spectra of injected protons 
with the life -- time $\tau_0\gg T_U$, $\sigma_{inj}=5$ for $E_{inj}=
10^2E_{GZK}$ (thin solid line) and $E_{inj}=10^8E_{GZK}$ (thick 
solid line). For $\tau_0\approx 0.125 T_U$, $E_{inj}=10^8E_{GZK}$, 
the same function is plotted for $\sigma_{inj}=5$ (dashed line) 
and $\sigma_{inj}=7$ (long dashed line). The observed fluxes are plotted 
by points. 
}
\end{figure}

\section{Some cosmological manifestations of decays of SHDM 
particles}

The injection of energy due to decays of SHDM particles during the 
"dark ages", at redshifts $10^3\geq z\geq 10$, leads also to 
interesting consequences some of which can be tested with available 
and/or future observations. Such consequences were recently 
discussed by Peebles, Seager and Hu (\cite{25}) in the framework 
of a simple toy model. Here we repeat this analysis using results 
obtained above. 

The decays of SHDM particles produce, among others, many high energy 
photons and electron -- positron pairs which, after reduction of their 
energy in electromagnetic cascades, are converted into $Ly-\alpha$ and 
$Ly-c$ photons with energies $E_\alpha=10.2$eV and $E_c=13.6$eV, 
respectively. The efficiency of such conversion is small due to high 
complexity of these cascades. To avoid many assumptions required for 
discussion of the final intensity and spectrum of photons at energy 
of interest we will assume, following \cite{25}, that the decays 
of SHDM particles lead to creation of $Ly-c$ and $Ly-\alpha$ photons 
with a rate  
\be
{d n_{ph}\over dt}\approx \varepsilon_{ph}{E_{inj}\over E_\alpha}
{n_0\over\tau_0}\approx \left({1+z\over 
1000}\right)^3\cdot 10^{-10}{\varepsilon_{ph}\over cm^3s },
\label{nph}
\ee  
where, for numerical estimates, we use $E_{inj}=10^8 E_{GZK}\approx 
10^{28}$ eV, the decay rate $n_0/\tau_0 = 10^{-46}cm^{-3}s^{-1}$, 
and $\varepsilon_{ph}\ll$ 1 characterizes the  unknown efficiency of energy 
transformation to $Ly-c$ and $Ly-\alpha$ photons. 

Comparing the rate of photons creation (\ref{nph}) with the rates 
discussed in \cite{25},
\be
{d n_\alpha\over dt}=\varepsilon_\alpha n_H H\approx 2.5\cdot 10^{-14}
\left({1+z\over 1000}\right)^{9/2}{\varepsilon_{\alpha}\over cm^3s },
\label{pb}
\ee
we see that, for $\varepsilon_\alpha\sim$ 1 -- 10 and correspondingly for 
\be
\varepsilon_{ph}\sim 3\cdot 10^{-4}\varepsilon_\alpha\sim 
10^{-4} - 10^{-3},
\label{lim}
\ee
the impact of discussed decays of SHDM particles effectively delays 
the recombination of hydrogen and leads to measurable distortions of 
the observed spectra of CMB fluctuations at angular wave numbers 
$l\geq 100 - 200$ (see detailed discussion in \cite{25}). 

For shorter life -- time of SHDM particles, $\tau_0\approx 0.125 T_U$, 
the decay rate at redshifts $z\geq$ 10 increases by about a 
factor of $10^3$ in comparison with (\ref{nph}) and wider range of 
$E_{inj}$ and $\varepsilon_{ph}$ can also be considered. The impact of
the factor $\omega_p$ omitted in (\ref{nph}) will also reinforce these 
estimates.
 
At smaller redshifts, $z\leq$ 500, generated Ly--c photons partly 
ionize neutral hydrogen. For small ionization degree of hydrogen, 
$x_H\leq$ 1, all Ly--c photons are rapidly absorbed and  
$x_H$ can be found from the equilibrium equation which describes 
the conservation of number of electrons and Ly--c photons together, 
\be
{d n_{ph}\over dt}=\alpha_{rec}^*n_e n_p =\alpha_{rec}^*\langle 
n_b\rangle^2 x_H^2
\label{xe}
\ee
where $\alpha_{rec}^*\approx 2\cdot 
10^{-13}(T/10^4 K)$ is the recombination coefficient for states with
the principle 
quantum number $n\geq$2, ~$T/10^4 K\approx 0.03[(1+z)/100]^2$ 
is the temperature of hydrogen under the condition of small 
ionization, $\langle n_b\rangle\approx 0.24(\Omega_bh^2/0.02)
[(1+z)/100]^3$ is the mean number density of baryons, and 
$n_p = n_e = x_H\langle n_b\rangle$. For simplicity, we neglected 
here the contribution of helium. 

As follows from Eqs. (\ref{nph}) and (\ref{xe}), the expected 
degree of hydrogen ionization is 
\be
x_H\approx \sqrt{\varepsilon_{ph}}\left({1+z\over 100}\right)^{-3/4}. 
\label{xh1}
\ee
For $\varepsilon_{ph}\geq 10^{-6}$, this degree is higher than the
degree of remaining hydrogen ionization
after  recombination, $x_H\sim 10^{-3}$. For shorter 
life -- time of  SHDM particles, $\tau_0\approx 0.125 T_U$,  
the decay rate at redshifts of interest grows by about a 
factor of $10^3$ and increases the ionization degree up to 
\be
x_H\sim \sqrt{10^3\varepsilon_{ph}}\left({1+z\over 100}\right)^{-3/4},
\label{xh2}
\ee
that again essentially extends the range of acceptable $E_{inj}$ 
and $\varepsilon_{ph}$. 

The considered growth of $x_H$, at redshifts $z\leq$ 500, does not 
increase significantly the optical depth for Thompson scattering, 
$\tau_T$, because 
\bedm
{d\tau_T\over dz}\propto x_H(1+z)^{-3/2}. 
\eedm
So, it does not amplify perturbations of the observed spectra 
of CMB fluctuations as compared with distortions generated at redshifts 
$z\sim$ 1000. But this growth essentially accelerates the creation 
of molecules $H_2$ and therefore formation and cooling of first galaxies. 
  
These results were obtained when the  extragalactic component 
of UHECR, discussed in Sec. III, dominates. These estimates can be also repeated 
for decays of more massive SHDM particles and usually discussed spectra 
of injected protons, with $\alpha = 1.5$ and $n_0/\tau_0\sim 10^{-46} 
cm^{-3} s^{-1}$.

\section{Summary and discussion}

Available information about possible properties of SHDM particles 
and, in particular, about their life -- time, energy and spectra 
of injected protons is now very limited, and, in fact, the analysis 
of  UHECR can be considered as an experimental test for  
particle interactions at ultra -- high energy. Results obtained 
in previous Sections show that under quite natural assumptions 
about the energy, life--time and spectrum of injected protons, a 
reasonable explanation of the observed fluxes of UHECR can be achieved. 
The CEL approximation used in this paper describes quite well the expected 
fluxes at $E\leq E_{GZK}$ but, for larger energies, the observed 
fluxes can be essentially distorted due to random character of 
energy losses above the GZK cutoff \cite{22}. 

For both log-normal and power spectra of injected protons, $S_{ln}$ 
(\ref{sgs}) and $S_{pw}$ (\ref{spc05}), with  $E_{inj}\geq 
10^5E_{GZK}$ and $\alpha\leq$1, and for suitable intensity of proton 
creation and life -- time of SHDM particles, our results are 
consistent with the observed fluxes for  $E_0\geq 0.06 E_{GZK}$. They 
demonstrate that, for $E_0\sim E_{GZK}$, the observed flux of UHECR 
can be mainly related to the extragalactic component. For such spectra 
of injected protons, the expected flux is found to be moderately 
sensitive to assumptions about the galactic sources, the energy of 
injection for  $E_{inj}\geq 10^5E_{GZK}$ and to values of $\alpha\leq$
0.6 and $\sigma_{inj}\leq$ 10 for power and log-normal spectra, 
respectively. In contrast, for the models with power spectra and  
$\alpha\geq$ 1 (\ref{spc1}), and for models with smaller energy of 
injection, $E_{inj}\sim 10^2 - 10^3 E_{GZK}$, the contribution of 
extragalactic component of UHECR is small, the galactic component 
dominates and the observed fluxes cannot be reproduced. 

\begin{figure}
\centering
\epsfxsize=8 cm
\epsfbox{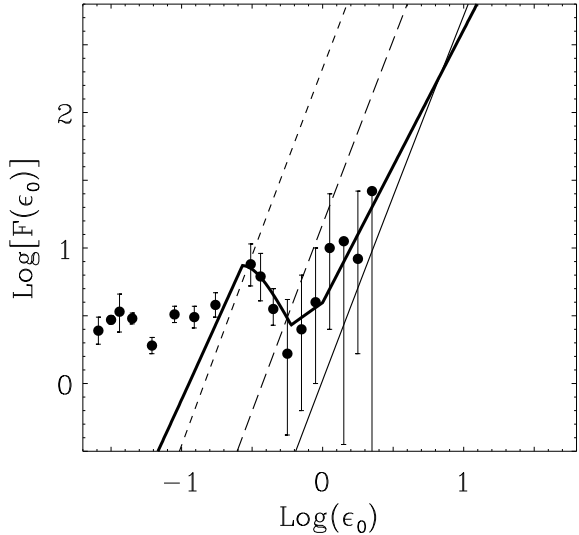}
\vspace{0.75cm}
\caption{The functions $F(\epsilon_0)$ (\ref{fe}) versus $\epsilon_0=
E/E_{GZK}$ are plotted for log-normal spectra of injected protons 
with the life -- time $\tau_0\gg T_U$, $\sigma_{inj}=5$ and $E_{inj}=
10^8E_{GZK}$. The extragalactic component is plotted by thick solid line, 
the galactic components are plotted for $\zeta_{gal}=10$ (solid line), 
$\zeta_{gal}=30$ (long dashed line) and $\zeta_{gal}=100$ (dashed 
line). 
}
\end{figure}

\subsection{Anisotropy of UHECRs}

For models under consideration with ultra -- high energy of injection, 
the observed fluxes for  $E_0\sim$ 0.1 -- 10$E_{GZK}$ are mainly related 
to the almost isotropic extragalactic component generated at redshifts 
$z\leq$ 0.1-- 0.2. At smaller and larger energies domination of the galactic 
component leads to an essential growth of anisotropy. This means that 
the anisotropy of angular distribution of observed UHECR is expected 
to be minimal for the energy range $E_0\sim$ 0.1 -- 10$E_{GZK}$. 

To illustrate this statement  we plot in Fig.6 the galactic and 
extragalactic fluxes separately for log--normal spectrum of injected 
protons for three values of $\zeta_{gal}$. 
As is seen from this Fig., for $\zeta_{gal}\leq$ 30 more isotropic 
extragalactic component dominates at $E\leq$ 1-- 10$E_{GZK}$ while 
for $\zeta_{gal}\geq$ 30 stronger anisotropy will be generated 
for all energies by the dominant galactic component. These variations of 
the anisotropy allow to test these versions of the Top -- Down models. 
The composition of extragalactic component of UHECR and, in particular, 
the possible contribution of high energy photons can also be considered 
as an important test of the model under discussion.

For simplicity and due to qualitative character of our analysis, 
we consider the CDM dominated flat cosmological model only. Evidently, 
these results can be recalculated in the same manner for other cosmological 
models and, in particular, for the most popular $\Lambda$CDM flat model. 
Of course, for such cosmological models some of the parameters used
here will be changed. However, even for such models the main qualitative 
results concerning the influence of the life -- time, mass and spectra 
of injected protons will remain. 

\subsection{Cosmological manifestations of decays of SHDM 
particles}

The discussed cosmological manifestations of the Top-Down model 
of UHECR generation provide an indirect test of this model. As is seen 
from evaluations given in Sec. IV for the high masses of SHDM particles 
and energy of injection $E_{inj}/E_{GZK}\sim10^8$, the effective 
delay of the cosmological recombination is possible for reasonable 
values of efficiency of creation of Layman photons. For models with 
decays of vortons, necklaces and other particles with $E_{inj}/E_{GZK}
\ll 10^8$ these cosmological manifestations are negligible.

The expected distortions of CMB fluctuations due to delayed  
recombination can be directly tested with the available modern 
balloon-born experiments (MAXIMA-1 \cite{26} and BOOMERANG \cite{27}) 
and future -- MAP and PLANCK satellite missions -- by measuring 
the CMB anisotropy and polarization power spectra. These 
problems will be discussed elsewhere. 

\subsection{Expected composition of UHECR and restrictions of the 
Top--Down models} 

Special problem is the composition of expected UHECRs. As is well 
known, decay products are dominated by pions, photons and neutrinos
while high energy protons make up only small part of these products. 
Therefore in the Top--Down models the observed flux of protons is  
accompanied by noticeable flux of high energy photons, electron - 
positron pairs and neutrinos. Probable energy losses of neutrinos are 
small \cite{6}, they are concentrated at $E\sim E_{inj}\gg E_{GZK}$
and could be responsible only for relatively rare observed events. 
But the comparison of expected and observed fluxes of high energy 
photons restricts some properties of SHDM particles and the Top--Down 
model as a whole (\cite{28}).   

The models under consideration predict the creation of high energy 
photons with $E\sim E_{inj}\gg E_{GZK}$. The possible evolution of 
such photons is quite uncertain as it depends upon unknown factors 
such as the extragalactic magnetic field and properties of radio 
background. For photons with $E_\gamma\gg E_{GZK}$, the free path, 
$D_\gamma\sim 100(1+z)^{-3}$Mpc, is, probably, defined by the process 
of double pair production in the CMB photons, $\gamma\gamma_b
\rightarrow e^+e^-e^+e^-$ \cite{6,29}, which leads to formation of 
less energy photons in electromagnetic cascades. For photons with 
$E_\gamma\sim E_{GZK}$, the free path, $D_\gamma\sim$ 1 Mpc$\cdot 
(E/E_{GZK})$ \cite{29}, is defined in main by their interaction with 
a badly known radio background. 

The flux and spectrum of the extragalactic component of photons differ 
from those expected for the protons. They depend upon the adopted 
life-time of the SHDM particles, properties of the electromagnetic 
cascades and radio background, and other factors. This complicated 
problem requires special detailed discussion. 

However, estimates of the photon free--path show that, for decays of 
the SHDM particles, the galactic component is dominated by photons 
with the spectrum of injection. The cumulative contribution of the 
photons is more then the cumulative contribution of the protons 
by a factor of $\sim\omega_p^{-1}\gg$ 1. This means that even for 
models under consideration the cumulative observed flux of UHECRs 
at energy $E\geq E_{GZK}$ is also dominated by galactic component 
of photons with $E\sim E_{inj}\gg E_{GZK}$. 
This conclusion follows from quite general arguments and, in fact, 
the discrimination of high energy photon component of UHECRs can be 
considered as the crucial test for the Top--Down models. 

Of course, the registration of such photons and the discrimination 
between photons with $E\gg E_{GZK}$ and protons with $E\sim E_{GZK}$ 
is special observational problem (see, e.g., \cite{30}). Thus, 
recently published restrictions \cite{3} of photon contribution relate 
to energy $E\leq 10^{20}$eV. 

Some factors can allow to suppress the expected flux of high energy 
photons. For example, this flux will be essentially suppressed if 
the photon free -- path within the Galaxy is $\sim$ 10 kpc or less. 
This means, however, that the radio background within the Galaxy must 
be more then that adopted in \cite{29} for the intergalactic space 
by a factor of $\sim$100 or more. 

Other important factor is the distribution of the 
SHDM particles within the halo of Galaxy. Thus, if the SHDM particles 
compose only relatively small fraction of the mean DM density of the 
universe then their possible segregation within halos also suppresses 
the isotropic flux of both galactic protons and photons. But such 
segregation increases the expected anisotropy of UHECRs with an 
essential excess of events from the centrum of Galaxy. Other versions 
of the improved Top -- Down models can be related to possible 
variations of composition of decay products and processes of proton 
creation at $E\gg E_{GZK}$ \cite{7,31}.

If sources of high energy protons are associated with observed galaxies 
as was recently discussed by Blanton, Blasi and Olinto (\cite{32}) for 
spectra of injected protons $S(E)\propto E^{-2}$ then restrictions 
related to the contribution of galactic components of high energy 
protons and photons become less important but more strong anisotropy 
of UHECRs is expected. It can be also expected that for such models 
with $S(E)\propto E^{-0.5}$ or log--normal spectrum of injected protons 
the resulting flux of UHECRs will be also similar to observed one. 

Available observational data do not yet allow to discriminate various 
explanations of creation of UHECRs with $E\geq E_{GZK}$ but further 
observations can give valuable information about particle interactions 
at ultra -- high energy, properties and spatial distribution of SHDM 
particles and some characteristics of the Galaxy and the universe at 
both small and high redshifts.  

\section*{Acknowledgment}
Authors are grateful to P. Blasi, P. Chardonet, M. Demianski, 
I. Novikov and I. Tkachev 
for discussions and help during the preparation of the paper.
This paper was supported in part by Danmarks Grundforskningsfond 
through its support for the establishment of the Theoretical 
Astrophysics Center, by grants RFBR 17625 and INTAS 97-1192.

\end{document}